# Polar Vortex Superstructure and Its Coupling with Correlated Electrons in Quasiperiodic Moiré Crystal

Si-yu Li[1,2]†, Zhongrui Wang[3]†, Yingzhuo Han[1]†, Shaoqing Xu[4]†, Zhiyue Xu[5]†, Yingbo Wang[1], Zhengwen Wang[1], Yucheng Xue[1], Aisheng Song[5], Kenji Watanabe[6], Takashi Taniguchi[7], Xueyun Wang[4], Tian-Bao Ma[5]*, Jiawang Hong[4]*, Hong-Jun Gao[1,2], Yuhang Jiang[3]*, Jinhai Mao[1]*

[1]School of Physical Sciences, University of Chinese Academy of Sciences, Beijing 100049, China

[2]Institute of Physics, Chinese Academy of Sciences, Beijing 100190, China

[3]College of Materials Science and Optoelectronic Technology, Center of Materials Science and Optoelectronics Engineering, University of Chinese Academy of Sciences, Beijing 100049, China

[4]School of Aerospace Engineering, Beijing Institute of Technology, Beijing 100081, China

[5]State Key Laboratory of Tribology in Advanced Equipment, Tsinghua University, Beijing 100084, China

[6]Research Center for Electronic and Optical Materials, National Institute for Materials Science, 1-1 Namiki, Tsukuba 305-0044, Japan

[7]Research Center for Materials Nanoarchitectonics, National Institute for Materials Science, 1-1 Namiki, Tsukuba 305-0044, Japan

*Corresponding author. Email: mtb@mail.tsinghua.edu.cn; hongjw@bit.edu.cn; yuhangjiang@ucas.ac.cn; jhmao@ucas.ac.cn

†These authors contributed equally to this work.

**Abstract:** Nanoscale polar structures are significant for understanding polarization processes in low-dimensional systems and hold potential for developing high-performance electronics. Here, we demonstrate a polar vortex superstructure arising from the reconstructed moiré patterns in twisted bilayer graphene aligned with hexagonal boron nitride. Scanning tunneling microscopy reveals spatially modulated charge polarization, while theoretical simulations indicate that the in-plane polarization field forms an array of polar vortices. Notably, this polar field is gate-tunable, exhibiting an unconventional gate-tunable polar sliding and screening process. Moreover, its interaction with electron correlations in twisted bilayer graphene leads to modulated correlated states. Our findings establish moiré pattern reconstruction as a powerful strategy for engineering nanoscale polar structures and emergent quantum phases in van der Waals materials.

Achieving precise and customizable modulation of properties in quantum materials has long been a research objective. Moiré superlattices (*1*) have garnered significant attention due to their ability to induce a variety of exotic quantum phases, including superconductivity (*2, 3*), quantum magnetism (*4, 5*) and unconventional ferroelectricity (*6-8*). Notably, the interaction between moiré patterns from adjacent interfaces can substantially reconfigure the structural and electronic properties of moiré heterostructures. For instance, super-moiré structures emerge in doubly aligned graphene-hexagonal boron nitride (hBN) systems (*9, 10*), and electric field-tunable superconductivity has been observed in twisted multilayer graphene (*11*). The tunable superlattice structures, periodicities, and orientation angles make the interplay between multiple moiré patterns a versatile tool for creating novel superstructures and quantum phases in a modular fashion (*12-14*). A prominent outcome of moiré-moiré interaction is moiré pattern reconstruction (MPR), characterized by pronounced strain fields with distinctive motifs (*12, 14-17*). Consequently, the associated electromechanical coupling effects are expected to induce charge polarization, significantly influencing the properties of the systems.

To date, charge polarization phenomena have been preliminarily explored in multi-moiré heterostructures, such as ferroelectric Chern insulators (*18*) and bistable superconductivity (*19*) in the twisted bilayer graphene (TBG) aligned with hBN. The mechanisms underlying these polarization states are under debate, with asymmetric moiré potentials and strain fields induced by MPR potentially serving as pivotal contributors. Specially, strain gradients-induced flexoelectricity has been detected in the graphene/hBN (*20*) and twisted graphene systems (*21-23*) despite graphene being conductive, reinforcing the theoretical and experimental research on polar metal (*24-26*). Given that local electric polarization strongly depends on the regional mechanical conditions, strain gradients can interact with the moiré ferroelectricity or independently generate polar superstructures (*27-30*). These polar superstructures open new avenues for nanoscale quantum state modulation by leveraging the sensitivity of correlated and topological properties of moiré systems to local charge density and electric fields. However, the detailed characteristics of electromechanical coupling, polar superstructures and their interplays with the quantum phases remain insufficiently understood, but are crucial for advancing high-performance electronics and polarization field-controlled devices.

In this article, we demonstrate a polar superstructure exhibiting gate tunability in a double moiré system, and investigate its interactions with the electron correlations through scanning tunneling microscopy/spectroscopy (STM/STS). Our devices are fabricated by aligning TBG with the hBN substrate, resulting in two distinct moiré patterns: the TBG and the G-hBN moiré (*31*). The strong interaction between these moiré patterns leads to significant MPR, generating inhomogeneous strain fields that substantially impact the electronic properties of TBG. Pronounced band bending is observed to evolve with the periodicity of the TBG moiré, providing spectroscopic evidence for charge polarization (*32-34*). We attribute the observed charge polarization to the electromechanical coupling effect induced by the strain gradients, namely flexoelectricity (*35*). Our theoretical calculations support this hypothesis and depict the polar superstructure as an array of polar vortices. Furthermore, by adjusting the back gate voltage, we observe modulations in polarization strength and the sliding of polar centers, underscoring its gate-tunable nature. Importantly, the electron correlations in TBG generally exhibits polarization-screening characteristics, and the correlated gaps show spatial variations related to the local polar conditions. The intriguing coupling between polar vortex superstructures and electron correlations is shown to enable manipulation of the positions of polar centers. Our work reveals that the interplay between different interfacial moiré patterns enables the realization of tunable polar vortex superstructures, and electromechanical coupling effects hold promise for effectively engineering novel quantum states.

**Moiré pattern reconstruction in the TBG-hBN**

The schematic diagram of TBG/hBN is illustrated in Fig. 1a, wherein $\theta_{TBG}$ denotes the twist angle between the two graphene layers, and $\theta_{hBN-G}$ is the twist angle between the bottom graphene and hBN substrate (Supplementary Fig. S1). The two twist angles cause two moiré patterns (the TBG moiré and the G-hBN moiré), directly confirmed by the STM image of the TBG-hBN device A and its fast Fourier transform with twelve Bragg peaks (Fig. 1b and the inset, more details in Supplementary Fig. S2). Due to the different energy scales at which the two moiré patterns impact the band structure, they can be separately visualized in-situ by adjusting the STM bias voltages (30). Specifically, the honeycomb-like G-hBN moiré with a wavelength ~ 15 nm ($\theta_{hBN-G} \sim 0°$) is prominent at higher bias voltages (Fig. 1c). In contrast, the triangular TBG moiré with a wavelength ~ 16.2 nm ($\theta_{TBG} \sim 0.87°$) is resolved at lower bias voltages (Fig. 1d). The two moirés showcase different morphological characteristics, which have been ascribed to their distinct structural relaxations for energy minimization (36, 37). For clarity, we superpose white hexagons to outline the G-hBN moiré and use black dots to label the AA sites of the TBG moiré, where the carbon atoms in the two graphene layers are well aligned.

Since the AA sites of the TBG moiré and the domain walls of the G-hBN moiré host high configuration energies, their overlap is minimized through lattice relaxation, at the expense of elastic energy (37, 38). Consequently, both moiré patterns undergo significant reconstruction. To quantitatively analyze the observed MPR, we extract the domain wall length $L$ and interior angle $\theta$ of the G-hBN moiré from the large-scale STM images of device A (Fig. 1e) and a single-moiré graphene/hBN device (Fig. 1f). The extracted lengths are presented in Fig. 1g and 1h, wherein the domain wall lengths in device A are considerably less uniform compared to those in the control device. The frequency distributions of $L$ and $\theta$ of both devices are shown in Fig. 1i and 1j. In the single-moiré G-hBN systems, these distributions closely follow a δ-profile, converging to constant values ($\theta$: 120˚; $L$ depends on $\theta_{hBN-G}$). While in device A, they are notably more dispersed. This contrast confirms that the interaction between the TBG and G-hBN moiré in device A induces additional structural relaxations, leading to more pronounced deformations of G-hBN moiré into irregular hexagons.

Next, we focus on the TBG moiré to further investigate the impacts of the MPR. Figure 2a shows the STM image of a TBG device without G-hBN moiré, where the TBG moiré patterns exhibit almost identical shape and size (moiré wavelength ~ 14.3 nm, twist angle ~ 0.98°). However, in device A (Fig. 2b), the AA sites of the TBG moiré are significantly reconstructed to exhibit irregular near-elliptical shapes oriented in various directions and lack translational symmetry, indicating varied local heterostrain conditions (Supplementary Fig. S3). To validate this scenario, height profiles across the AA sites enclosed by the dashed circles in both devices are extracted along three directions (Fig. 2c and d). The curves in Fig. 2d show clear non-overlap, sharply contrasting with those from the single-moiré TBG device (Fig. 2c), confirming that the MPR can effectively enhance structural relaxation and generate spatially modulated strain fields.

**The impacts of MPR on the electronic properties of TBG**

The intriguing MPR in moiré heterostructures have attracted much attention for their essential roles in determining the band structures and ground states (*12, 14, 15, 36-38*). One noteworthy outcome of this phenomenon is the electromechanical coupling induced charge polarization, particularly flexoelectricity (*20-22*), which becomes pronounced at the nanoscale in inhomogeneous structural deformations (*35*). The flexoelectric effect would lead to a spatially modulated charge polarization with complex polar landscapes that differ dramatically from the uniform electrical polarization within the bulk of conventional ferroelectric materials (*27-30, 39-41*). In STM experiments, the presence of charge polarization is evidenced by the distinct band energies at different spatial positions, as the polarized charge modifies the local Coulomb potential and finally leads to band bending (*32-34*). Figures 2e-f present the *dI/dV* spectra measured at the AA and AB sites of TBG moiré in device A and a controlled TBG device, with flat bands highlighted by black and blue arrows (*42*). Beyond the modulation of spectral intensity, the flat bands at the AA and AB sites in the single-moiré TBG device exhibit nearly identical energies (Fig. 2e). In stark contrast, the flat bands show energy shifts in device A (Fig. 2f). These results preliminarily reveal the existence of charge polarization in the TBG/hBN, which is closely linked to the observed MPR.

**Characterizations of the charge polarization by theoretical simulations and STM**

Based on our recent work (*43*), we employ molecular dynamics (MD) to corroborate the MPR and charge polarization in a TBG-hBN system with periodic boundary conditions (Supplementary Fig. S4). The simulation successfully captures the main features of the observed MPR in device A, including the pronounced distortions of the two sets of moiré patterns (Fig. 2g, where the AA sites of TBG moiré are highlighted by black dots). Furthermore, the strain vector field (Fig. 2h), the strength maps of in-plane strain ($\varepsilon_{ij}$) and related strain gradients ($\varepsilon_{ij,k}$) in the TBG can be obtained, since the balanced positions of C atoms and their in-plane displacements after relaxation have been determined (Supplementary Fig. S5-S7). Strain gradients can give rise to flexoelectric polarization, which has been experimentally observed by piezoresponse force microscopy (PFM) in other twisted graphene systems (*21-23*). Building on this, we calculate the magnitude of the in-plane polarization field ($|P_{//}|$) in the top graphene layer, as directly probed by the STM tip (Fig. 2i). The maximum values of $|P_{//}|$ predominantly surround the AA sites, forming a triangular-lattice-like distribution. To elucidate the spatial evolutions of this polarization field, we present the local polar vectors (the black arrows in Fig. 2j) in the region marked by the purple hexagon in Fig. 2g. Notably, the directions of polar vectors gradually deflect, forming an array of vortex-like structures with the AA sites as the centers (more details in Supplementary Fig. S8).

Similar complex polar textures have previously been observed in other moiré heterostructures (such as twisted BaTiO$_3$ and MoS$_2$) and strained epitaxial thin films, but only by scanning transmission electron microscopy (*27, 44, 45*).

To further explicate the charge polarization, we calculate the polarization charge density in the top graphene, and find that the positive polarization charges mainly reside at the AA sites (marked by black dots in Fig. 3a; the polarization charge density in bottom graphene shown in Supplementary Fig. S9). The varying polarization charge density causes spatial modulation of the Coulomb potential (Supplementary Fig. S11), leading to spatial band bending (*33, 46*). We investigate this effect in device A by tracking the energy evolution of the flat bands. Figure 3c presents representative *dI/dV* spectra, detailing the spatial evolution of the flat bands (*42*). Besides energy shifts, these spectra also reveal spatial variations in the energy spacing between the two flat bands (marked by arrows in Fig. 3c), underscoring the significant influence of mechanical relaxation on the intrinsic band structures of TBG (*43*). The color-scale map of the line-cut *dI/dV* spectra along the direction indicated by the blue arrow in Fig.3b further illustrates band bending and energy spacing modulation (Fig. 3d). For clarity, the corresponding peak positions of the valence flat band (VFB) and conduction flat band (CFB) are replotted in Fig. 3e as purple and green dots respectively, with solid lines representing fitted curves and dashed lines indicating their average values. Our results demonstrate that the observed band bending follows the TBG moiré, qualitatively compatible with the calculated flexoelectric charge polarization in Fig. 3a. By the equation $P = (\varepsilon_r - 1)\varepsilon_0\, \delta V/\delta d$, we calculate the in-plane polarization field $P_{//}$, where $\varepsilon_r = 4.5$ is the relative permittivity of graphene, $\varepsilon_0$ is vacuum electric permittivity, $\delta V/\delta d$ is the derivative of the flat-band peak energy shift with respect to position. The calculated result from the spatial evolution of VFB is presented in Fig. 3f, with a maximum value of approximately 0.04 $\mu C \cdot cm^{-2}$, comparable to the reported polarization strengths in graphene-based heterostructures or other 2D materials (*6, 25, 47*).

Due to the modulation of the flat-band bending, the *dI/dV* intensity maps would undergo a spatial contrast inversion with energy variation. At relatively low energies, the flat-band peaks are located at the AA sites, but shift toward the transition regions as the energy approaches the Fermi level. This contrast inversion is demonstrated by the *dI/dV* maps at -60 mV and -4 mV (Fig. 3g and h, with dots marking the AA sites), solidly validating that charge polarization is not confined to a specific direction but is instead widely distributed, closely following the TBG moiré. Additional experimental results demonstrating MPR and band bending in various configurations, obtained from other TBG-hBN samples with different moiré wavelengths and relative orientations compared to device A, are presented in Supplementary Fig. S12 and S13. These findings underscore the high tunability of charge polarization and the associated polarization fields in TBG-hBN (Table S1).

**The interaction between the electron correlation and charge polarization**

The detailed interplay between the charge polarization and correlated electronic states in moiré heterostructures remains an intriguing but open question. To explore this, we focus on our TBG/hBN device B, where the TBG moiré wavelength is approximately 14.7 nm ($\theta_{TBG} \sim 0.96°$,

$\theta_{\text{G-hBN}} \sim 0°$), similar to that of magic-angle graphene (Supplementary Fig. S14). Figure 4a shows *dI/dV* spectra measured at the center of an AA site, where the cascade of electronic transitions (marked by black arrows) and band splitting are evidently visible. When the filling factors $v \geq 4$ ($v \leq -4$), flat bands are fully occupied (empty) with electrons and can be trivially described by non-interacting models. While at the filling factors between -4 to 4, the Coulomb interactions turn to dominate over the kinetic energy of the electrons, further breaking the degeneracy of spin and valley degrees of freedom and splitting flat bands into sub-bands (*31, 42*). Near each integer fillings, the spin/valley flavor polarization is reset to cause the reconfigurations of the low-energy excitations, which are directly visualized by the spectroscopic features highlighted by the arrows in Fig. 4a.

Interestingly, we observe band bending throughout the filling process, regardless of whether the system is dominated by single-particle physics (Fig. 4b and 4g) or electron correlations (Fig. 4c-f). The charge polarization modulates the strongly correlated electronic properties, as evidenced by the spatial modulation of the correlated gaps at $v \approx -3$ and -2.5 (Fig. 4f and Supplementary Fig. S15). The spatial evolution of the flat-band peak energy at various fillings is summarized in Fig. 4h, where the black arrows indicate regions with the highest positive polarization charge density. As the filling factor varies, the fine features of band bending evolve, accompanied by shifts in the spatial positions of effective positive polarization centers. This suggests that charge polarization in TBG-hBN is highly tunable via gating (Fig. 4i), with a non-monotonic dependence on gate voltage, reflecting its intricate interplay with carrier density, external electric fields, and electron correlations. The drift of positive polarization centers indicates modifications in both the spatial distribution of polarization charge and the direction of the polar field (Fig. 4j and k). Yet, no discernible changes in topographic features are observed (Supplementary Fig. S16). Moreover, in device C where $\theta_{\text{TBG}}$ largely deviates from the magic angle (Supplementary Fig. S17), no polar sliding is observed, underscoring the critical role of electron correlations which can give rise to selective filling of electrons at different spatial locations (*33*). These findings unveil that the mechanism of the charge polarization in TBG-hBN is primarily driven by electronic effects (*48, 49*).

**Outlook**

Recently, efforts to engineer charge polarization in 2D materials through stacking, sliding, and twisting have been explored, yielding promising progresses such as novel 2D ferroelectric patterns intertwined with flexoelectricity (*27-29*). Our work unveils the polar vortex superstructures originating from the spatially varied strain field in TBG-hBN, where the coupling with the charge polarization and electron correlations allows for gate-tunable control over these polar vortex states. The celebrated advantage of multi-moiré systems, exemplified by TBG-hBN, is that MPR can be readily tailored by tuning the wavelengths, lattice structures, and relative orientations of modular moiré patterns (*11-13, 15, 17*). Thus, diverse polar superstructures can be realized, introducing new possibilities to achieve polarization-dependent physical phenomena. We believe our work provides a valuable approach to construct novel polar microstructures potentially with non-trivial topologies and exotic quantum phases, and would advance future research such as the unconventional ferroelectricity in 2D materials.

**References and Notes**

1. D. M. Kennes *et al.*, Moiré heterostructures as a condensed-matter quantum simulator. *Nature Physics* **17**, 155-163 (2021).

2. Y. Cao *et al.*, Unconventional superconductivity in magic-angle graphene superlattices. *Nature* **556**, 43-50 (2018).

3. O. Can *et al.*, High-temperature topological superconductivity in twisted double-layer copper oxides. *Nature Physics* **17**, 519-524 (2021).

4. C. L. Tschirhart *et al.*, Imaging orbital ferromagnetism in a moiré Chern insulator. *Science* **372**, 1323-1327 (2021).

5. E. Anderson *et al.*, Programming correlated magnetic states with gate-controlled moiré geometry. *Science* **381**, 325-330 (2023).

6. Z. Zheng *et al.*, Unconventional ferroelectricity in moiré heterostructures. *Nature* **588**, 71-76 (2020).

7. A. Weston *et al.*, Interfacial ferroelectricity in marginally twisted 2D semiconductors. *Nature Nanotechnology* **17**, 390-395 (2022).

8. K. Yasuda, X. Wang, K. Watanabe, T. Taniguchi, P. Jarillo-Herrero, Stacking-engineered ferroelectricity in bilayer boron nitride. *Science* **372**, 1458-1462 (2021).

9. X. Sun *et al.*, Correlated states in doubly-aligned hBN/graphene/hBN heterostructures. *Nature Communications* **12**, 7196 (2021).

10. Z. Wang *et al.*, Composite super-moiré lattices in double-aligned graphene heterostructures. *Science Advances* **5**, eaay8897 (2019).

11. Z. Hao *et al.*, Electric field–tunable superconductivity in alternating-twist magic-angle trilayer graphene. *Science* **371**, 1133-1138 (2021).

12. S. Turkel *et al.*, Orderly disorder in magic-angle twisted trilayer graphene. *Science* **376**, 193-199 (2022).

13. A. Uri *et al.*, Superconductivity and strong interactions in a tunable moiré quasicrystal. *Nature* **620**, 762-767 (2023).

14. X. Lai *et al.*, Imaging Self-aligned Moiré Crystals and Quasicrystals in Magic-angle Bilayer Graphene on hBN Heterostructures. arXiv:2311.07819 (2024).

15. I. M. Craig *et al.*, Local atomic stacking and symmetry in twisted graphene trilayers. *Nature Materials* **23**, 323-330 (2024).

16. L. Wang *et al.*, New Generation of Moiré Superlattices in Doubly Aligned hBN/Graphene/hBN Heterostructures. *Nano Letters* **19**, 2371-2376 (2019).

17. N. R. Finney *et al.*, Tunable crystal symmetry in graphene–boron nitride heterostructures

with coexisting moiré superlattices. *Nature Nanotechnology* **14**, 1029-1034 (2019).

18. M. Chen *et al.*, Selective and quasi-continuous switching of ferroelectric Chern insulator devices for neuromorphic computing. *Nature Nanotechnology* **19**, 962-969 (2024).

19. D. R. Klein *et al.*, Electrical switching of a bistable moiré superconductor. *Nature Nanotechnology* **18**, 331-335 (2023).

20. X. Jiang *et al.*, Unravelling the electromechanical coupling in a graphene/bulk h-BN heterostructure. *Nanoscale* **14**, 15869-15874 (2022).

21. Y. Li *et al.*, Unraveling Strain Gradient Induced Electromechanical Coupling in Twisted Double Bilayer Graphene Moiré Superlattices. *Advanced Materials* **33**, 2105879 (2021).

22. L. J. McGilly *et al.*, Visualization of moiré superlattices. *Nature Nanotechnology* **15**, 580-584 (2020).

23. H. Zhang *et al.*, Layer-Dependent Electromechanical Response in Twisted Graphene Moiré Superlattices. *ACS Nano* **18**, 17570-17577 (2024).

24. P. W. Anderson, E. I. Blount, Symmetry Considerations on Martensitic Transformations: "Ferroelectric" Metals? *Physical Review Letters* **14**, 217-219 (1965).

25. Z. Fei *et al.*, Ferroelectric switching of a two-dimensional metal. *Nature* **560**, 336-339 (2018).

26. J. Shi *et al.*, Revealing a distortive polar order buried in the Fermi sea. *Science Advances* **10**, eadn0929 (2024).

27. G. Sánchez-Santolino *et al.*, A 2D ferroelectric vortex pattern in twisted $BaTiO_3$ freestanding layers. *Nature* **626**, 529-534 (2024).

28. A. N. Morozovska *et al.*, Flexoinduced ferroelectricity in low-dimensional transition metal dichalcogenides. *Physical Review B* **102**, 075417 (2020).

29. S. Wan *et al.*, Intertwined Flexoelectricity and Stacking Ferroelectricity in Marginally Twisted hBN Moiré Superlattice. *Advanced Materials* **36**, 2410563 (2024).

30. G. Catalan *et al.*, Flexoelectric rotation of polarization in ferroelectric thin films. *Nature Materials* **10**, 963-967 (2011).

31. M. Oh *et al.*, Evidence for unconventional superconductivity in twisted bilayer graphene. *Nature* **600**, 240-245 (2021).

32. K. Chang *et al.*, Discovery of robust in-plane ferroelectricity in atomic-thick SnTe. *Science* **353**, 274-278 (2016).

33. S.-y. Li *et al.*, Imaging topological and correlated insulating states in twisted monolayer-bilayer graphene. *Nature Communications* **13**, 4225 (2022).

34. C. J. Butler *et al.*, Mapping polarization induced surface band bending on the Rashba semiconductor BiTeI. *Nature Communications* **5**, 4066 (2014).

35. B. Wang, Y. Gu, S. Zhang, L.-Q. Chen, Flexoelectricity in solids: Progress, challenges, and perspectives. *Progress in Materials Science* **106**, 100570 (2019).

36. N. N. T. Nam, M. Koshino, Lattice relaxation and energy band modulation in twisted bilayer graphene. *Physical Review B* **96**, 075311 (2017).

37. P. San-Jose, A. Gutiérrez-Rubio, M. Sturla, F. Guinea, Spontaneous strains and gap in graphene on boron nitride. *Physical Review B* **90**, 075428 (2014).

38. L. Huder *et al.*, Electronic Spectrum of Twisted Graphene Layers under Heterostrain. *Physical Review Letters* **120**, 156405 (2018).

39. Y. L. Tang *et al.*, Observation of a periodic array of flux-closure quadrants in strained ferroelectric $PbTiO_3$ films. *Science* **348**, 547-551 (2015).

40. C.-L. Jia, K. W. Urban, M. Alexe, D. Hesse, I. Vrejoiu, Direct Observation of Continuous Electric Dipole Rotation in Flux-Closure Domains in Ferroelectric $Pb(Zr,Ti)O_3$. *Science* **331**, 1420-1423 (2011).

41. S. Chen *et al.*, Recent Progress on Topological Structures in Ferroic Thin Films and Heterostructures. *Advanced Materials* **33**, 2000857 (2021).

42. D. Wong *et al.*, Cascade of electronic transitions in magic-angle twisted bilayer graphene. *Nature* **582**, 198-202 (2020).

43. S.-y. Li *et al.*, Quasiperiodic Moiré Reconstruction and Modulation of Electronic Properties in Twisted Bilayer Graphene Aligned with Hexagonal Boron Nitride. *Physical Review Letters* **133**, 196401 (2024).

44. C. S. Tsang *et al.*, Polar and quasicrystal vortex observed in twisted-bilayer molybdenum disulfide. *Science* **386**, 198-205 (2024).

45. D. Du, J. Hu, J. K. Kawasaki, Strain and strain gradient engineering in membranes of quantum materials. *Applied Physics Letters* **122**, 170501 (2023).

46. J. Mao *et al.*, Realization of a tunable artificial atom at a supercritically charged vacancy in graphene. *Nature Physics* **12**, 545-549 (2016).

47. J. Deng *et al.*, Ferroelectricity in an Antiferromagnetic Vanadium trichloride monolayer. arXiv:2404.13513 (2024).

48. M. Springolo, M. Royo, M. Stengel, Direct and Converse Flexoelectricity in Two-Dimensional Materials. *Physical Review Letters* **127**, 216801 (2021).

49. J. Xu, Z. Zhang, P. Cui, Coexistence and interplay of pseudomagnetism and flexoelectricity in few-layer rippled graphene. *npj Quantum Mater.* **9**, 102 (2024).

50. W. B. Donald *et al.*, A second-generation reactive empirical bond order (REBO) potential energy expression for hydrocarbons. *Journal of Physics: Condensed Matter* **14**, 783 (2002).

51. C. Sevik, A. Kinaci, J. B. Haskins, T. Çağın, Influence of disorder on thermal transport properties of boron nitride nanostructures. *Physical Review B* **86**, 075403 (2012).

52. I. Leven, T. Maaravi, I. Azuri, L. Kronik, O. Hod, Interlayer Potential for Graphene/h-BN Heterostructures. *Journal of Chemical Theory and Computation* **12**, 2896-2905 (2016).

53. T. Maaravi, I. Leven, I. Azuri, L. Kronik, O. Hod, Interlayer Potential for Homogeneous Graphene and Hexagonal Boron Nitride Systems: Reparametrization for Many-Body Dispersion Effects. *The Journal of Physical Chemistry C* **121**, 22826-22835 (2017).

54. L. Shu, X. Wei, T. Pang, X. Yao, and C. Wang, Symmetry of flexoelectric coefficients in crystalline medium, *Journal of Applied Physics* **110**, 104106 (2011).

55. L. Wang *et al.*, Flexoelectronics of centrosymmetric semiconductors. *Nature Nanotechnology* **15**, 661-667 (2020).

56. A. T. Pierce *et al.*, Unconventional sequence of correlated Chern insulators in magic-angle twisted bilayer graphene. *Nature Physics* **17**, 1210-1215 (2021).

57. K. P. Nuckolls *et al.*, Strongly correlated Chern insulators in magic-angle twisted bilayer graphene. *Nature* **588**, 610-615 (2020).

58. S. Grover *et al.*, Chern mosaic and Berry-curvature magnetism in magic-angle graphene. *Nature Physics* **18**, 885-892 (2022).

59. A. A. Shabana. Computational Continuum Mechanics. *Cambridge University Press*, New York, 2008.

60. J. Bonet, R. D. Wood. Nonlinear Continuum Mechanics for Finite Element Analysis. *Cambridge University Press*, Cambridge, 1997.

61. S. Zhu, J. A. Stroscio, and T. Li, Programmable Extreme Pseudomagnetic Fields in Graphene by a Uniaxial Stretch, *Physical Review Letters* **115**, 245501 (2015).

62. D.-H. Kang *et al.*, Pseudo-magnetic field-induced slow carrier dynamics in periodically strained graphene, *Nature Communications* **12**, 5087 (2021).

**Acknowledgments:** We sincerely thank Prof. Jianming Lu, Prof. Jiamin Xue and Prof. Jian Kang for their valuable discussions, as well as Xiaonan Xu for assistance with data analysis.

**Funding:**

the National Key R&D Program of China (Grant No. 2019YFA0307800, and 2021YFA1400303)

the National Natural Science Foundation of China (Grants No. 11974347, No. 12474477, No. 12074377, No. 12172047, No. 52225502, No. 52305200, No. 52363033 and No. 61888102)

the China Postdoctoral Science Foundation with Certificate No. 2024M753465

the Postdoctoral Fellowship Program (Grade C) of China Postdoctoral Science Foundation with Grant No. GZC20241893

Fundamental Research Funds for the Central Universities

the EMEXT Element Strategy Initiative to Form Core Research Center through Grant No. JPMXP0112101001

**Author contributions:**

Conceptualization: J.M., Y.J.

Methodology: J.M., J.H., T.M, A.S., Z.X., S.L.

Investigation: S.L., Z.W., Y.H., S.X.

Visualization: S.L., Z.W., Y.H., S.X.

Funding acquisition: J.M., Y.J., H.G., J.H., T.M., S.L.

Project administration: J.M., Y.J., H.G., J.H.

Supervision: J.M., Y.J., H.G., J.H.

Writing – original draft: S.L., J.M., Y.J.

Writing – review & editing: All authors read and approved of the final version of the manuscript

**Competing interests:** The authors declare no competing interests.

**Data and materials availability:** All data are available in the main text or the supplementary materials.

## Supplementary Materials

Materials and Methods

Supplementary Text

Figs. S1 to S17

Tables S1

References (*50–62*)

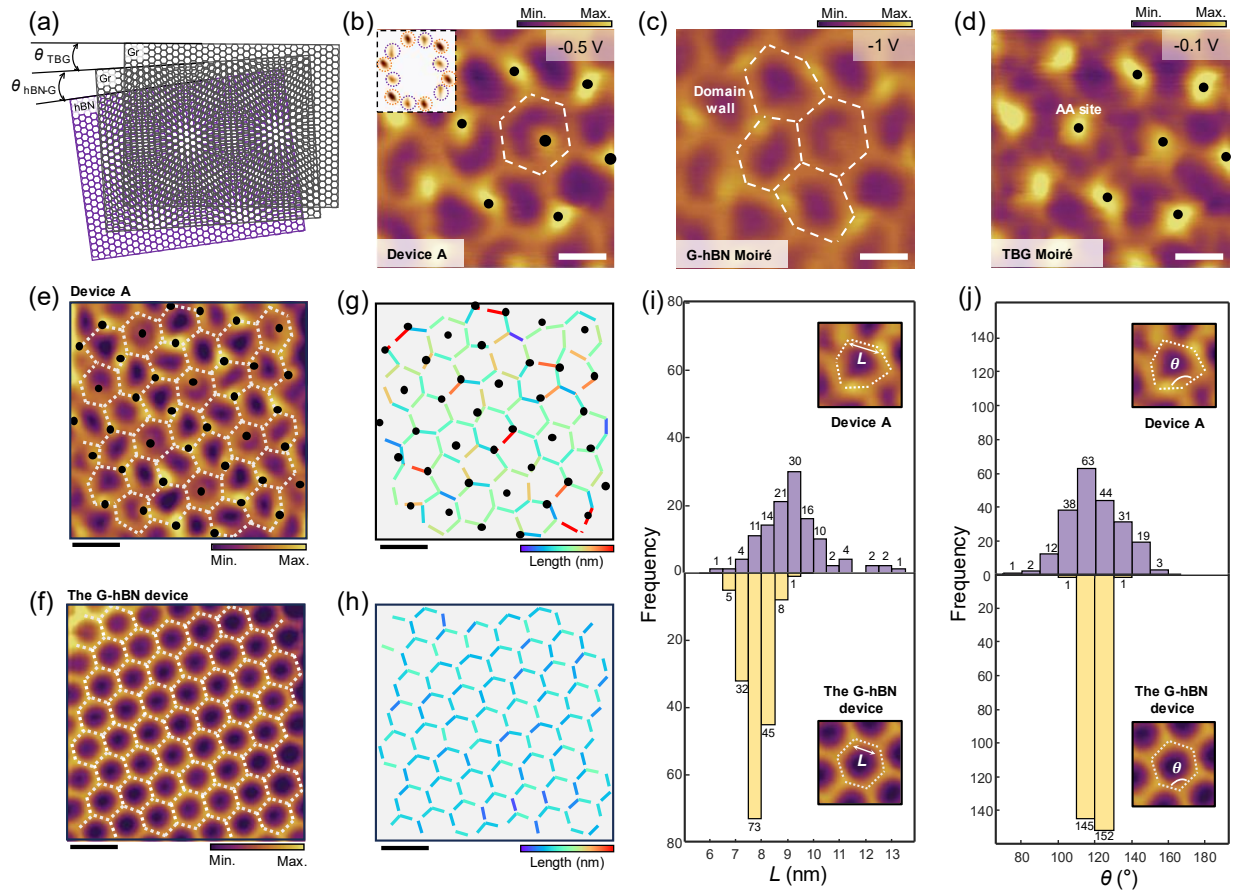

**Fig. 1 The two distinct interfacial moiré and the reconstructed G-hBN moiré patterns in TBG/hBN.** (a) The schematic diagram of TBG/hBN, where $\theta_{TBG}$ is the twist angle between the two layers of graphene, and $\theta_{hBN-G}$ is the twist angle between the bottom graphene and hBN substrate. (b-d) The STM topographies of the moiré patterns in device A with tunneling current $I$ = 30 pA and bias voltage $V_b$ = −0.5 V (b), -1 V (c) and -0.1 V (d), respectively. The inset in (b) is its FFT and shows two sets of reciprocal lattices from the TBG and the G-hBN moiré, which are labelled with the black dots and white dashed lines respectively. (e-f) The large scale STM images of device A ($I$ = 30 pA and $V_b$ = −1 V), and the single-moiré G-hBN device (monolayer graphene aligned with hBN, $I$ = 20 pA and $V_b$ = −0.4 V). (g-h) The sketch maps of the side lengths $L$ of the G-hBN moiré patterns in (e) and (f). Color bars: 6 nm to 12 nm. (i-j) The frequency histograms of the side lengths $L$ (i) and the interior angles $\theta$ (j) of the G-hBN moiré patterns of device A (upper panels) and the G-hBN device (lower panels). The insets depict the definitions of $L$ and $\theta$ in the zoom-in topographies. The scalebars are 10 nm for (b-d) and 20 nm for (e-h).

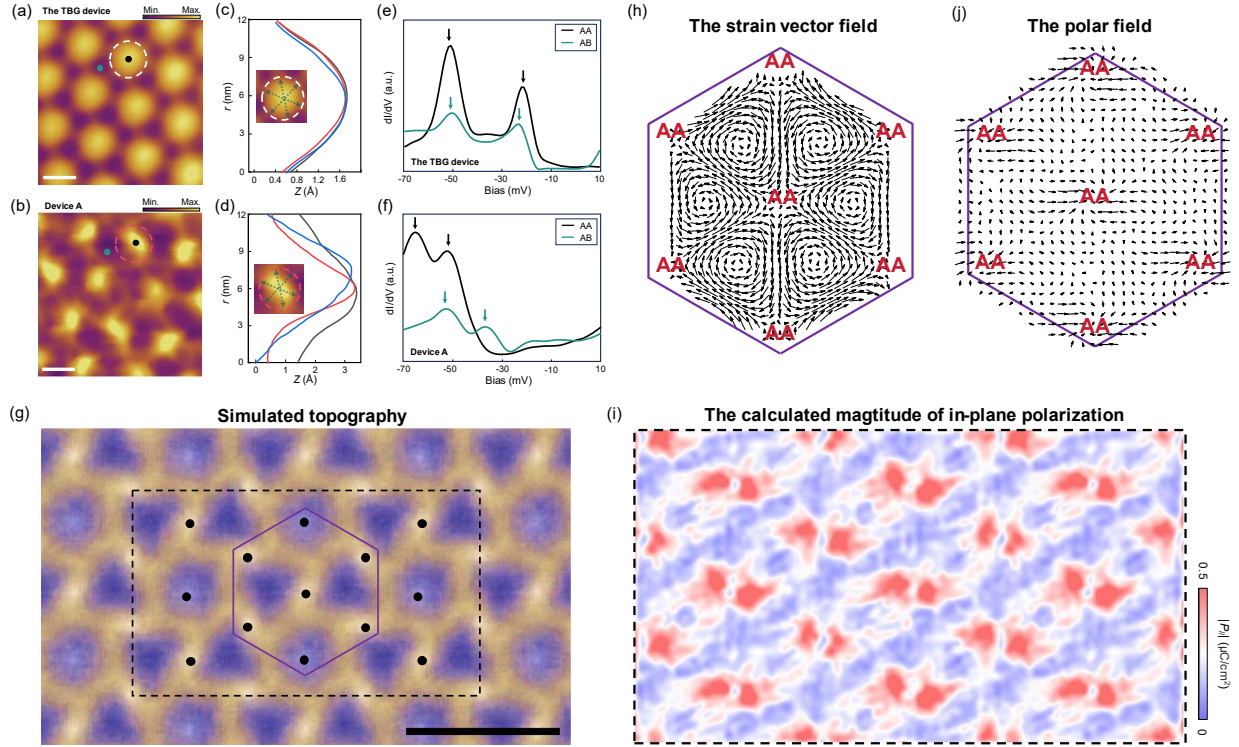

**Fig. 2 The reconstructed TBG moiré patterns, flat-band energy shift and polarization field in TBG/hBN.** (a-b) The STM topography of the controlled TBG device misaligned with the hBN substrate with $I$ = 30 pA, $V_b$ = −1 V (a), and device A with $I$ = 300 pA, $V_b$ = −0.1 V (b). (c-d) The height profile in three directions of the AA site labelled by the white dashed circle in (a), and labelled by the red dashed circle in (b). (e-f) The $dI/dV$ spectra at the AA and AB region (marked by the black and blue dots) in the controlled TBG device ($I$ = 400 pA and $V_b$ = −0.1 V) and device A. The black and blue arrows highlight the flat bands. (g) The MD simulated topography of TBG/hBN, where the black dots mark the AA sites. Scale bar, 30 nm. (h) The calculated strain vector field ($e_x$, $e_y$) where $e_x = \varepsilon_{xx} - \varepsilon_{yy}$ and $e_y = -(\varepsilon_{yx} + \varepsilon_{xy})$ in the region marked by the purple hexagon in (g). (i) The magnitude map of the calculated in-plane polarization field in the region highlighted by the black rectangle in (g). (j) The calculated vorticity map of polar field in the top graphene in the region highlighted by the black rectangle in (g), where the white arrows represent the local vector components of the calculated in-plane polarization field.

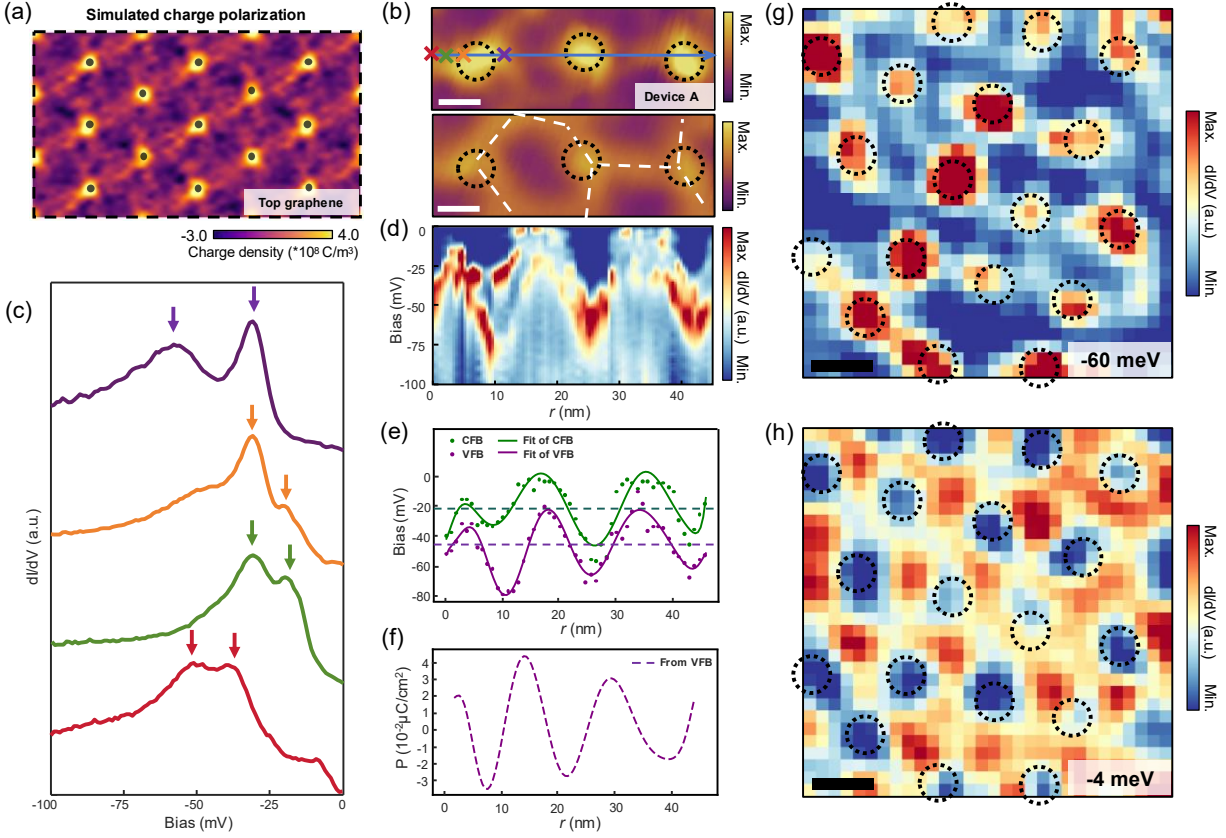

**Fig. 3 Flat-band bending and electromechanical coupling induced charge polarization in TBG/hBN.** (a) The simulated charge polarization in the top graphene of TBG/hBN in the region highlighted by the black rectangle in Fig. 2(g), where the black dots mark the AA sites. (b) STM images of distorted moiré patterns. Upper panel, topography for TBG moiré. The AA sites are labelled by black circles. Lower panel, topography for G-hBN moiré, highlighted by white dashed lines. (c) $dI/dV$ spectra collected at the crosses in (b) with $I$ = 150 pA, $V_b$ = −0.1 V. (d) Intensity plot of $dI/dV$ spectra taken along the blue arrow in (c). (e) Spatial dependence of the peak energies of conduction flat band (CFB, green) and valence flat band (VFB, purple), with raw (fitted) data represented by dots (lines) and their average peak energies labelled by the dashed lines. The peak energies of CFB and VFB are extracted from the local maximum of the $dI/dV$ spectra in (d). (f) Spatial dependence of the electric polarization field $P$ derived from the spatial evolutions of peak energies of VFB. (g-h) $dI/dV$ maps at (g) -60 meV and (h) -4 meV, depicting evolutions of local density of states and nearly ordered charge polarizations.

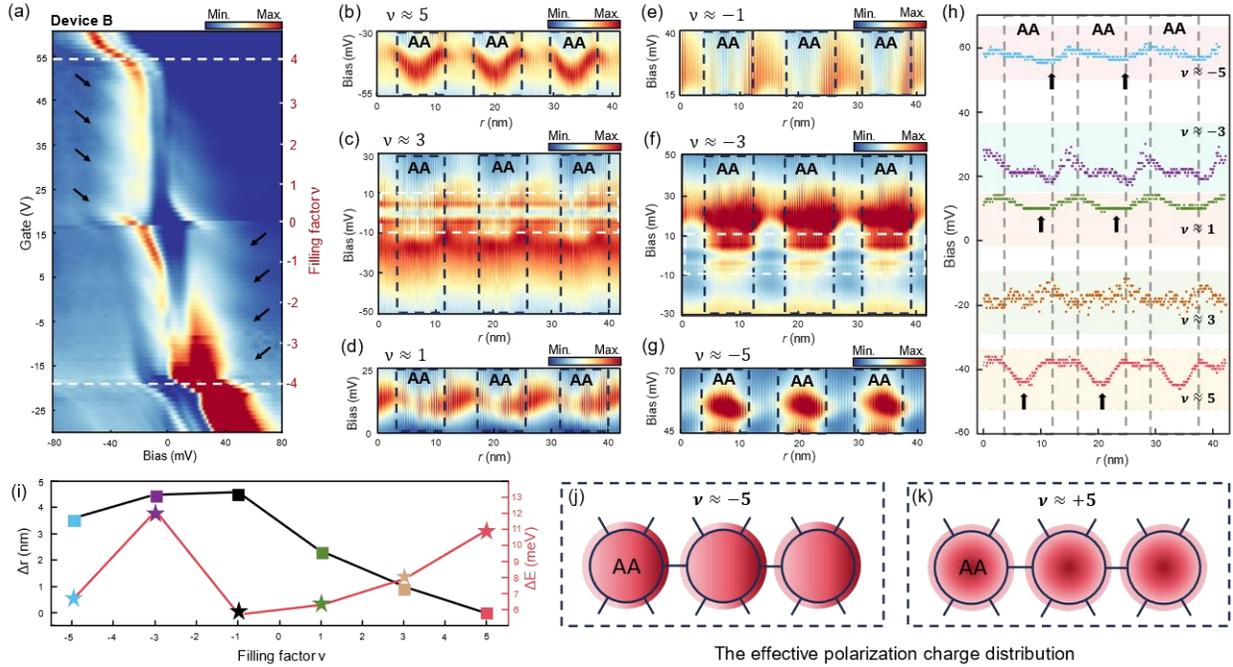

**Fig. 4 The gate-tunable charge polarization and its interactions with the correlated states in TBG/hBN.** (a) Back-gate (filling factor) dependence of $dI/dV$ spectra measured at an AA site in device B ($I$ = 300 pA, $V_b$ = −80 mV). The cascade of correlated electronic transitions is highlighted by black arrows. The fully electron (hole) doping, referring to $v$ = +4 (-4), are labelled by white dashed lines. (b-g) Intensity plot for spatial dependence of $dI/dV$ spectra at (b) $v \approx$ +5, (c) $v \approx$ +3, (d) $v \approx$ +1, (e) $v \approx$ -1, (f) $v \approx$ -3, (g) $v \approx$ -5. The AA sites are labelled by black dashed squares, revealing band bending at different filling factors. The correlated gaps at $v \approx$ +3 and $v \approx$ -3 are highlighted by white dashed squares. (h) Spatial dependence of the peak energies of flat bands at $v \approx$ +5 (red dots), $v \approx$ +3 (orange dots), $v \approx$ +1 (green dots), $v \approx$ -3 (purple dots), $v \approx$ -5 (blue dots). The AA sites are labelled by gray dashed squares, and the polarization centers are highlighted by the black arrows, indicating their spatial shift induced by doping. (i) The modulations of polarization with filling factors. Black line: the drifts of polar centers away from the center of AA sites. Red line: the degrees of band bending. (j, k) The schematic diagrams of effective polarization charge distribution at two different filling factors $v \approx$ -5 (i) and $v \approx$ +5 (j), where darker red indicates a higher density of polarized positive charges.